\documentclass[prb,aps,superscriptaddress,reprint]{revtex4-2}

\usepackage{hyperref}
\hypersetup{colorlinks=true, linkcolor=blue, citecolor=blue, urlcolor=blue}
\usepackage{graphicx}
\usepackage{amsmath}
\usepackage{siunitx}
\usepackage{mathptmx}
\usepackage{xcolor}
\usepackage[normalem]{ulem}
\begin{document}

\title{Magnetic proximity beyond the interface: Layer-sequence-dependent exchange in WSe$_2$/CrI$_3$ heterostructures}

\author{Natalie Kuhn}
\author{Marc Sch\"utte}
\author{Joachim Emming}
\author{Felix Hellenkamp}
\author{Leo Dreyer}
\author{Frank Volmer}
\affiliation{2nd Institute of Physics and JARA-FIT, RWTH Aachen University, 52074 Aachen, Germany}
\author{Kenji Watanabe}
\affiliation{Research Center for Electronic and Optical Materials, National Institute for Materials Science, 1-1 Namiki, Tsukuba 305-0044, Japan}
\author{Takashi Taniguchi}
\affiliation{Research Center for Materials Nanoarchitectonics, National Institute for Materials Science,  1-1 Namiki, Tsukuba 305-0044, Japan}
\author{Lutz Waldecker} 
\affiliation{2nd Institute of Physics and JARA-FIT, RWTH Aachen University, 52074 Aachen, Germany}
\author{Christoph Stampfer} 
\affiliation{2nd Institute of Physics and JARA-FIT, RWTH Aachen University, 52074 Aachen, Germany}\affiliation{Peter Gr\"unberg Institute (PGI-9), Forschungszentrum J\"ulich, 52425 J\"ulich, Germany}
\author{Bernd Beschoten} 
\email{bernd.beschoten@physik.rwth-aachen.de}
\affiliation{2nd Institute of Physics and JARA-FIT, RWTH Aachen University, 52074 Aachen, Germany}


\begin{abstract}
Magnetic proximity effects in transition metal dichalcogenide (TMD)/2D-magnet heterostructures are often described in terms of the magnetization of the interfacial magnetic layer. Here, we show that this picture is sufficient to describe the neutral-exciton valley splitting of the TMD, but incomplete for its hole-doped electronic structure. We study a gate-tunable van der Waals heterostructure composed of monolayer WSe$_2$ in direct contact with trilayer CrI$_3$. By combining spatially and helicity-resolved reflection-contrast spectroscopy with magnetic circular dichroism, we identify two distinct magnetic domains in the trilayer CrI$_3$: one with purely antiferromagnetic interlayer coupling and one with a mixed antiferromagnetic~(AFM)/ferromagnetic~(FM) coupling. In both domains, the sign of the neutral-exciton valley splitting of WSe$_2$ switches with the magnetization reversal of the adjacent CrI$_3$ layer. In contrast, helicity-resolved positive-trion spectroscopy of WSe$_2$ reveals a qualitatively different response. While the AFM domain shows a strongly helicity-selective trion resonance, the mixed AFM/FM domain exhibits trion resonances in both helicities. This behavior disappears when all CrI$_3$ layers are aligned in a parallel magnetization direction. Our results demonstrate that the positive-trion response is sensitive to the magnetic configuration of CrI$_3$ beyond the directly adjacent layer, revealing a magnetic proximity response that cannot be described solely by the interfacial magnetization.
\end{abstract}

\maketitle

\section{Introduction}
Van der Waals (vdW) heterostructures provide a versatile platform for proximitizing distinct two-dimensional (2D) materials and thereby tailoring their electronic, optical, and magnetic properties via interfacial proximity coupling~\cite{Geim2013Jul, Gmitra2015Oct, Wang2015Jan, Schaibley2016, Zhao2017Aug, Lee2018Mar, Mak2019Nov, Sierra2021Aug, Choi2022Jul, Volmer2023Aug, Zollner2024Dec}. A prominent example is the magnetic  proximity effect in vdW heterostructures composed of monolayer transition metal dichalcogenides (TMDs) and 2D-magnets, where exchange interactions induce magnetism in an otherwise non-magnetic semiconductor~\cite{Zhao2017Aug, Norden2019Sep, Livio2020, Zollner2019Aug, Zhong2020}. 

Monolayer TMDs such as WSe$_2$ are direct-bandgap semiconductors with two energetically degenerate valleys at the K and K$'$ points of the Brillouin zone. At the band edge, these valley states differ not only in crystal momentum, but also in the internal orbital character of their Bloch wave functions. Together with the strong spin-orbit coupling, which ties the spin polarization to the valley index, this valley-dependent orbital structure gives rise to circular optical selection rules making the two valleys  selectively addressable by circularly polarized light~\cite{Mak2012, Xiao2012, Xu2014, Yang2015Oct, Volmer2017Jun, Ersfeld2019Jun, Ersfeld2020May}. This unique coupling makes TMDs promising for spintronic~\cite{Sierra2021Aug} and valleytronic~\cite{Schaibley2016} applications, where the valley degree of freedom serves as an information carrier. Beyond optical excitation, an external magnetic field lifts the valley degeneracy via the valley Zeeman effect~\cite{MacNeill2015, Srivastava2015, Aivazian2015, Wang2015}, leading to valley splittings on the order of \SI{0.2}{meV/T}~\cite{MacNeill2015}.

\begin{figure*}[tb]
    \includegraphics[width=\textwidth]{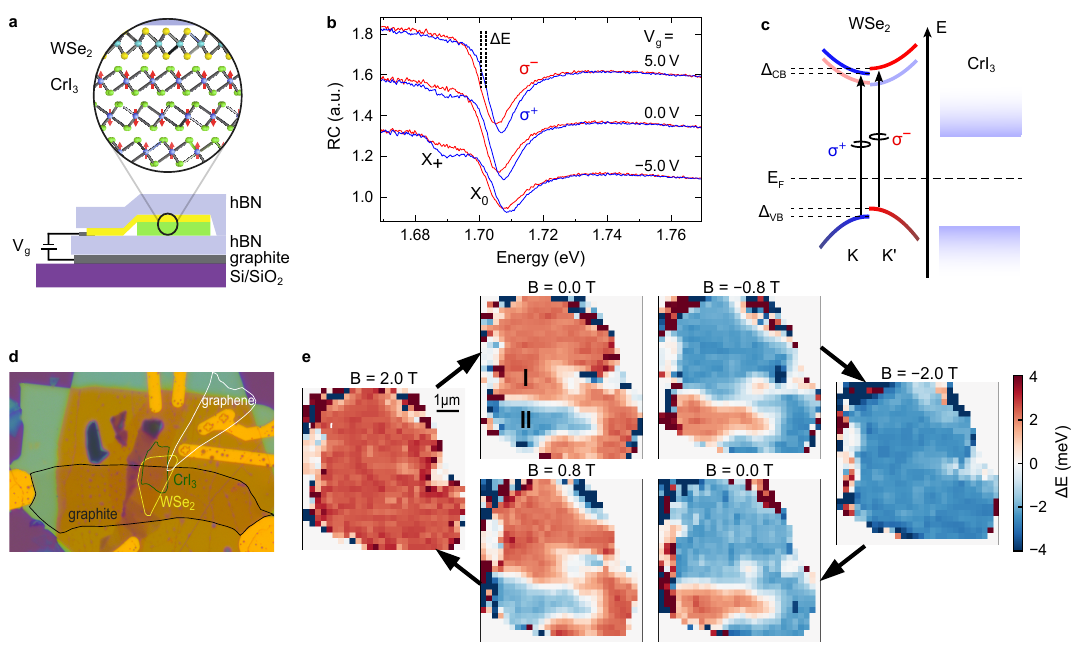}
    \caption[width=\textwidth]{\textbf{Device structure and spatially-resolved magnetic proximity effects.} \textbf{a} Schematic cross-section of a gate-tunable WSe$_2$/trilayer~CrI$_3$ heterostructure, fully encapsulated in hBN. A few-layer graphite backgate and a graphene contact to WSe$_2$ enable electrostatic doping.    
    \textbf{b} Helicity resolved reflection contrast at representative back gate voltages measured in region I, see panel (e). The positively charged trion (X$_+$) is especially prominent at negative gate voltages and vanishes at V$_g=$\SI{5}{V}. The energy difference between the two light helicities of the neutral exciton (X$_0$) defines the proximity-induced valley splitting $\Delta E$ of the WSe$_2$. 
    \textbf{c} Schematic band alignment of the WSe$_2$/CrI$_3$ heterostructure. The magnetic proximity effect lifts the valley degeneracy of WSe$_2$ while the staggered type-II band alignment between WSe$_2$ and CrI$_3$ allows the Fermi level to be tuned into the WSe$_2$ valence band. 
    \textbf{d} Optical microscope image of the device. The outlines indicate the monolayer WSe$_2$, trilayer CrI$_3$, graphite, and graphene flakes.
    \textbf{e} Spatial maps of the exciton valley splitting $\Delta E \approx \Delta_\mathrm{CB}-\Delta_\mathrm{VB}$ taken at selected magnetic fields during magnetic field down- and up-sweeps. At large positive and negative fields, the splitting is spatially uniform. At intermediate fields, two regions with distinct switching behavior appear: region I exhibits a single sign reversal of $\Delta E$ per sweep, whereas region II exhibits three sign reversals.}
    \label{fig:fig1}
\end{figure*}

With the discovery of 2D-magnets~\cite{Huang2017, Gong2017} and their integration into vdW-heterostructures, a new route has emerged to lift the valley degeneracy in monolayer TMDs via the magnetic proximity effect~\cite{Zhong2017, Livio2020}.
An example is monolayer WSe$_2$ proximity-coupled to few-layer CrI$_3$ ~\cite{Zhong2017}, where exchange interactions induce valley splittings in WSe$_2$ on the order of 2 to \SI{3}{meV}~\cite{Zhong2017, Seyler2018, Li2020, Zhong2020}.
Few-layer CrI$_3$ exhibits intralayer ferromagnetic order with an out-of-plane magnetic moment, while adjacent layers are typically coupled antiferromagnetically at zero magnetic field (see red arrows in the zoom-in of Fig.~\ref{fig:fig1}a). Owing to the relatively weak interlayer coupling, the magnetic configuration in CrI$_3$ is highly tunable. An external out-of-plane magnetic field can induce layer-by-layer magnetization switching~\cite{Huang2017, Song2018, Zhong2020}, while external pressure can stabilize ferromagnetic coupling even at zero magnetic field~\cite{Song2019_pressure_CrI3, Li2019_pressure_CrI3}.
Layer-resolved studies of WSe$_2$/CrI$_3$~heterostructures have shown that the sign of the proximity-induced valley splitting is predominantly determined by the magnetization of the CrI$_3$ layer adjacent to WSe$_2$. At the same time, the magnitude of the valley splitting can depend on the full magnetic configuration of the multilayer stack~\cite{Zhong2020}.
However, experiments have reported valley splittings with opposite signs and qualitatively distinct magnetic-field hysteresis behaviors within a single WSe$_2$/CrI$_3$ device~\cite{Zhong2017, Zhong2020, Seyler2018}. These observations suggest that the proximity effect may not be fully captured by an interfacial picture alone and raise the open question of how the magnetic layer sequence in few-layer CrI$_3$ influences the proximity-induced electronic structure in adjacent TMDs.

Here, we investigate a gate-tunable WSe$_2$/CrI$_3$-heterostructure using optical reflectance spectroscopy~\cite{Tebbe2023Dec}. We identify two spatially separated magnetic domains in the trilayer CrI$_3$ at zero magnetic field: a domain with purely antiferromagnetic interlayer coupling, hereafter referred to as the AFM domain, and a domain with mixed antiferromagnetic (AFM) and ferromagnetic (FM) interlayer coupling. By combining the proximity-induced valley splitting in WSe$_2$ with the magnetic circular dichroism of CrI$_3$ as a function of an external magnetic field, we resolve the layer-dependent magnetic configurations of the vdW heterostructure.
Moreover, electrostatic gating enables hole doping of WSe$_2$ and the formation of positive trions. The trion resonance serves as a probe of the spin- and valley-polarized valence-band states and resident holes.
We observe distinctly different trion responses in the two magnetic domains. In the AFM domain, the trion resonance appears predominantly in the helicity corresponding to the higher energy exciton resonance. In contrast, in the mixed AFM/FM domain, the trion resonance is visible in both helicities with comparable amplitudes. Once all CrI$_3$ layers in the mixed AFM/FM domain are switched into a parallel alignment by an external magnetic field, the trion response becomes similar to that of the AFM domain. These observations indicate that the trion response, and thus the spin- and valley-polarized valence-band states and resident holes in WSe$_2$, are influenced by the magnetic configuration of the CrI$_3$ trilayer beyond the adjacent layer. This sensitivity is qualitatively different from that of the neutral-exciton valley splitting, whose sign follows the interfacial CrI$_3$ magnetization.

\section{Results}

\begin{figure*}[tb]
    \includegraphics[width=\textwidth]{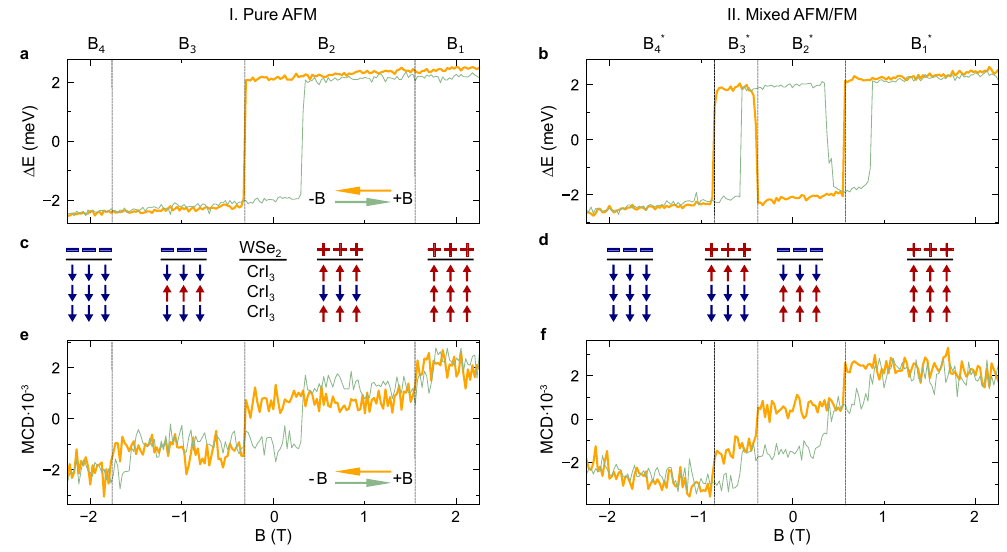}
    \caption[width=\textwidth]{\textbf{Reconstruction of layer-resolved magnetic configurations.} \textbf{a,b} Magnetic field hysteresis of the proximity-induced neutral-exciton valley splitting $\Delta E$ in WSe$_2$ for the AFM domain (region I) and the mixed AFM/FM domain (region II), respectively. Orange and green traces correspond to the two magnetic-field sweep directions, as indicated by the arrows. Region I shows a single reversal, whereas region II shows three sign reversals.
    \textbf{c,d} Reconstructed layer-resolved magnetic configurations of the heterostructure for the two regions. Plus and minus signs denote the sign of the WSe$_2$ valley splitting, while red upward and blue downward arrows indicate the out-of-plane magnetization direction of the individual CrI$_3$ layers. The configurations are labeled according to the down-sweep sequence B$_1$--B$_4$ for region I and B$_1^{*}$--B$_4^{*}$ for region II; the up-sweep follows the reversed sequence with hysteretically shifted switching fields. 
    \textbf{e,f} Magnetic circular dichroism (MCD) signal of CrI$_3$ for regions I and II. The MCD signal is proportional to the total CrI$_3$ magnetization and exhibits four discrete levels, corresponding to the  net magnetizations of the magnetic trilayer. The vertical dashed lines mark the down-sweep switching fields used to define the magnetic-field regimes in \textbf{c,d}.
    }
    \label{fig:fig2}
\end{figure*}

The device consists of a monolayer WSe$_2$ proximity-coupled to a trilayer CrI$_3$ flake and fully encapsulated in hexagonal boron nitride (hBN).
Both layer numbers were determined by optical contrast measurements~\cite{Uslu2024,Uslu2025Dec}. 
A few-layer graphite flake serves as a bottom  gate. Electrical contacts to WSe$_2$ are realized via a graphene layer. 
A schematic cross-section and an optical image are shown in Figs.~\ref{fig:fig1}a and ~\ref{fig:fig1}d, respectively. 

Figure~\ref{fig:fig1}b shows helicity-resolved reflection-contrast (RC) spectra recorded at gate voltages of $V_\text{g}=$~\SI{-5}{V}, \SI{0}{V}, and \SI{5}{V} at a temperature of 5~K. All spectra exhibit a resonance at energies around $1.70$ to \SI{1.71}{eV}, which we assign to the neutral exciton X$_{0}$ in WSe$_2$. The energy shift between the exciton resonances, measured by the excitation and probing with either $\sigma^{+}$ or $\sigma^{-}$ polarized light, stems from the proximity-induced valley splitting, which is approximately \SI{2}{meV} at zero magnetic field and zero gate voltage. 
Towards negative gate voltages ($V_\text{g}=$\SI{-5}{V}), the amplitude of the X$_{0}$ resonance is decreasing, whereas the resonance of the positively charged trion X$_{+}$ at around \SI{1.69}{eV} is increasing in amplitude. The appearance of the trion X$_{+}$, which consists of an electron-hole pair bound to an additional hole from the opposite valley, indicates gate-induced hole doping of WSe$_2$.
Notably, the trion exhibits a pronounced helicity dependence, with a significantly larger amplitude in the circular polarization corresponding to the higher energy exciton (blue lines in Fig.~\ref{fig:fig1}b), indicating that both valleys in the valence band are differently occupied. No negatively charged trion X$_{-}$ is observed at any gate voltage. The presence of X$_{+}$ and the absence of X$_{-}$ indicates that the Fermi level can be tuned into the valence band but not into the conduction band of WSe$_2$, consistent with a staggered (type-II) band alignment of the heterobilayer (see Fig.~\ref{fig:fig1}c). This band alignment is in agreement with previous reports~\cite{Zhong2017} and supported by DFT/GW calculations~\cite{Heissenbuettel2021}. The measured neutral-exciton valley splitting corresponds to the difference between the valley-dependent conduction- and valence-band shifts, $\Delta E \approx \Delta_\mathrm{CB}-\Delta_\mathrm{VB}$ (Fig.~\ref{fig:fig1}c), assuming valley-independent exciton binding energies (see Methods section on details about the fitting procedure and our sign convention of $\Delta E$).

We next use the proximity-induced exciton splitting in WSe$_2$ as a direct measure of the valley splitting and as a local probe of the magnetic proximity effect. Using this approach, we map the spatial variation of $\Delta E$ across the heterostructure as a function of an out-of-plane magnetic field (see methods section C for more details). Spatial maps of the valley-splitting taken at representative magnetic fields during the magnetic hysteresis are shown in Fig.~\ref{fig:fig1}e. At $\pm$\SI{2}{T} the valley splitting is spatially uniform across the heterostructure, reaching values of approximately $\pm$\SI{2}{meV}. Upon sweeping the magnetic field from \SI{+2}{T} to \SI{0}{T}, two distinct domains are seen. While most of the sample retains a positive valley splitting (red color, region I), a region in the lower-left corner (region II) exhibits a valley splitting of opposite sign (blue color). Upon further sweeping the magnetic field to \SI{-0.8}{T}, both regions reverse their valley splitting and at \SI{-2}{T} the splitting is again uniform, now with a value of approximately \SI{-2}{meV} (blue color).     
Reversing the sweep direction from \SI{-2}{T} to \SI{+2}{T} (lower branch in Fig.~\ref{fig:fig1}e) yields a similar overall behavior. While region I exhibits a single sign reversal of the valley splitting per sweep, region II displays three sign reversals within the same field range. The corresponding local hysteresis curves for both sweep directions are shown in Figs.~\ref{fig:fig2}a and \ref{fig:fig2}b. Since the valley splitting in WSe$_2$ results from the interface exchange field from CrI$_3$, these contrasting hysteresis behaviors indicate that the two regions experience different interfacial magnetization sequences. 

\subsection{Layer-resolved valley splitting and interlayer magnetic order}

We investigate the distinct valley-splitting behavior observed in regions I and II by analyzing the magnetic-field hysteresis of the proximity-induced exciton splitting in WSe$_2$, together with the magnetic circular dichroism (MCD) signal of CrI$_3$, which probes the total magnetization of the latter (see methods section for more details). The full hysteresis curves for both magnetic-field sweep directions are shown in Figs.~\ref{fig:fig2}a,b for the exciton valley splitting and in Figs.~\ref{fig:fig2}e,f for the corresponding CrI$_3$ MCD signal. In the following, we use the down-sweep to label the magnetic-field regimes B$_1$--B$_4$ and B$_1^{*}$--B$_4^{*}$ (orange curve), while the up-sweep (green curve) confirms the same layer-resolved configurations with hysteretically shifted switching fields.

We first discuss region I. The MCD hysteresis curve in Fig.~\ref{fig:fig2}e exhibits four discrete MCD levels with three switchings, defining four magnetic-field regimes B$_{1}$--B$_{4}$ along the down-sweep. Since the MCD signal is proportional to the total magnetization of CrI$_3$, these discrete MCD levels correspond to distinct magnetic configurations of the trilayer. The switching fields and overall hysteresis shape are consistent with previously reported MCD hysteresis curves of anti-ferromagnetically coupled CrI$_3$~\cite{Zhong2020, Song2018, Song2019_pressure_CrI3}. Fig.~\ref{fig:fig2}a depicts the exciton valley splitting and shows a splitting of around $\SI{+2}{meV}$ at $B=\SI{2.25}{T}$, which reverses its sign to about $\SI{-2}{meV}$ during the down sweep. Notably, only a single sign reversal of the valley splitting is observed. This sign reversal coincides with the transition between the magnetic field regimes B$_2$ and B$_3$ identified in the MCD hysteresis. The up-sweep shows the corresponding reversed sequence, confirming that region~I maintains a purely AFM-type switching sequence for both sweep directions.

We now consider region II. The MCD hysteresis curve in Fig.~\ref{fig:fig2}f also shows four MCD levels, defining the magnetic-field regimes B$_{1}^{*}$--B$_{4}^{*}$ along the down sweep. In contrast to region I, however, the switching fields associated with the highest and lowest MCD levels are significantly reduced, occurring at $B\sim \SI{0.5}{T}$ and $\SI{-0.9}{T}$, respectively, whereas the corresponding switching fields in case of region I are above $\SI{\pm 1.5}{T}$. Similar reductions of the magnetic switching fields have previously been reported for few-layer CrI$_3$ systems exhibiting mixed AFM/FM coupling, which were attributed to different stacking orders of the CrI$_3$ layers~\cite{Li2019_pressure_CrI3, Song2019_pressure_CrI3}. The proximity-induced valley splitting in region II, shown in Fig.~\ref{fig:fig2}b, exhibits a qualitatively different behavior from that in region I. Instead of a single sign reversal, three sign reversals of the valley splitting are observed during the magnetic-field down-sweep. Notably, the magnetic fields at which these sign reversals occur coincide with the transitions between the MCD levels in Fig.~\ref{fig:fig2}f.

To identify the layer-resolved magnetic configurations underlying these observations, we introduce a minimal model for the CrI$_3$ trilayer. Each layer contributes a magnetic moment $m_i = \pm 1$ depending on the orientation of its out-of-plane magnetization. The total magnetization of the trilayer is therefore given by $M = m_1 + m_2 + m_3$, yielding $M = \pm 1, \pm 3$, consistent with the four discrete MCD levels seen in Figs.~\ref{fig:fig2}e,f. While the MCD signal provides a measure of the total magnetization $M$, the sign of the proximity-induced valley splitting (Figs.~\ref{fig:fig2}a,b) directly reflects the magnetization of the CrI$_3$ layer adjacent to WSe$_2$ ($m_1$). The combination of these two observables, together with the characteristic hysteresis sequence observed in both sweep directions, allows us to infer the layer-resolved magnetic configurations. The schematics of the reconstructed layer configurations for both domains are illustrated in Figs.~\ref{fig:fig2}c,d. Here, the plus and minus signs correspond to a positive and a negative valley splitting of WSe$_2$, respectively, while the red upward and blue downward arrows represent the out-of-plane magnetization of each individual CrI$_3$ layer. 
It is important to note that the combination of the MCD signal and the sign of the neutral-exciton valley splitting does generally not uniquely determine the magnetization direction of each CrI$_3$ layer. While the MCD signal provides the total magnetization and the exciton splitting predominantly reflects the interfacial magnetization, additional information from the characteristic magnetic switching sequences and the known interlayer coupling behavior is required to reconstruct the complete magnetic configurations. The assignments shown in Figs. 2c and 2d are therefore based on the combined experimental observations and the established magnetic properties of trilayer CrI$_3$.

In region I, the transition from B$_1$ ($M=+3$) to B$_2$ ($M=+1$) does not change the sign of the valley splitting, indicating that the interfacial layer remains polarized upward ($m_1=+1$). Consequently, one of the remaining layers must reverse its magnetization. The observed MCD hysteresis, which matches that of antiferromagnetically coupled trilayer CrI$_3$, favors a configuration in which adjacent layers are antiparallel. This constraint uniquely selects the configuration to ($+1,-1,+1$), corresponding to a reversal of the middle layer. Applying the same reasoning to the transition to B$_3$ ($M=-1$) yields the configuration ($-1,+1,-1$). Finally, at large enough magnetic fields the magnetic moments of all layers are aligned to the external magnetic field, yielding ($-1,-1,-1$).

In contrast, region II exhibits a qualitatively different switching sequence. The transition from B$_{1}^{*}$ ($M=+3$) to B$_{2}^{*}$ ($M=+1$) is accompanied by a sign reversal of the valley splitting, indicating that the interfacial CrI$_3$ layer switches its magnetization. The only configuration consistent with $M=+1$ is therefore ($-1$,$+1$,$+1$). Extending this reasoning to the subsequent transition to B$_{3}^{*}$ ($M=-1$) leads to the configuration ($+1$,$-1$,$-1$) and again to ($-1$,$-1$,$-1$) at large magnetic fields in region B$_{4}^{*}$.

\begin{figure*}[tb]
    \includegraphics[width=0.8\textwidth]{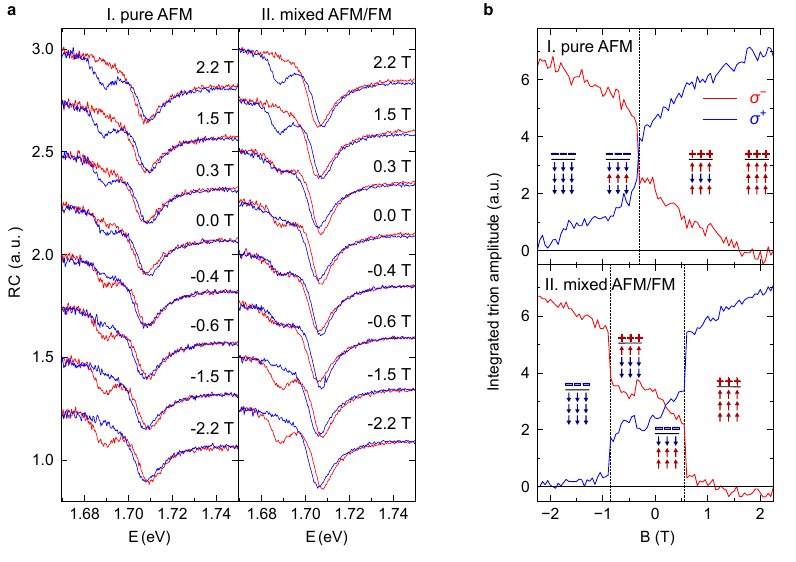}
    \caption[width=\textwidth]{\textbf{Different trion response in the two distinct magnetic domains.}
    \textbf{a} Helicity-resolved RC spectra of WSe$_2$ at selected magnetic fields during the magnetic field down-sweep of  the AFM domain and the mixed AFM/FM domain, measured at a gate voltage of V$_g=$\SI{-5}{V}. In the AFM domain, the positive trion X$_{+}$ is only dominant in one of the two light helicities, whereas it is almost absent in the other. The helicity, in which the feature can be observed, switches when the interfacial CrI$_3$ layer reverses. Instead, in the mixed domain, the trion becomes visible in both helicities in the intermediate magnetic field regime (see e.g. curve at \SI{-0.4}{T}).  
    \textbf{b} Integrated trion amplitudes for $\sigma^{+}$ and $\sigma^{-}$ polarization as a function of magnetic field. In the AFM domain, the two helicities remain strongly asymmetric throughout the sweep. In the mixed AFM/FM domain, the trion amplitudes become comparable between approximately $\SI{0.5}{T}$ to $\SI{-0.9}{T}$, coinciding with the mixed magnetization sequence of the CrI$_3$ trilayer.
    }
    \label{fig:fig3}
\end{figure*}

This reconstruction reveals purely AFM interlayer couplings in region I and mixed AFM/FM couplings between adjacent layers in CrI$_3$ in region II. Accordingly, we refer to region I as the AFM domain and to region II as the mixed AFM/FM domain. In the trilayer system, the WSe$_2$ serves as a probe of the interfacial CrI$_3$ magnetization, enabling us to determine which of the remaining layers align ferromagnetically. This provides direct access to the layer-resolved magnetic order and demonstrates the presence of a partially ferromagnetic alignment in CrI$_3$.

\subsection{Trion spectroscopy of layer-resolved magnetic configurations}

After reconstructing the layer-resolved magnetic configuration of the CrI$_3$ trilayer, we next examine whether the RC signals of trions in WSe$_2$ are also controlled solely by the interfacial CrI$_3$ layer, as it is the case for the neutral excitons in both the AFM- and mixed AFM/FM-domains of the heterostructure. To this end, we use the positively charged trion $X_+$ as a charge-sensitive optical probe for the electronic states near the valence band maximum of WSe$_2$. In contrast to the neutral exciton, whose energy splitting primarily tracks the proximity-induced valley splitting, the appearance of the $X_+$ resonance depends also on the availability of resident holes in the spin-valley-locked valence-band states. Helicity-resolved trion spectroscopy therefore provides access both to the valley-dependent hole population and to changes of the valence band alignment in the proximitized WSe$_2$.

Figure~\ref{fig:fig3}a shows helicity-resolved RC spectra measured at V$_\text{g}=$\SI{-5}{V} during the magnetic-field down-sweep for the AFM and the mixed AFM/FM domains. In the AFM domain (left panel), the trion resonance is strongly helicity-selective. At all magnetic fields, the trion appears predominantly in the one helicity that corresponds to the blueshifted neutral exciton resonance (blue curves for $B\geq\SI{0}{T}$, red curves for $B\leq\SI{-0.4}{T}$). Accordingly, upon sweeping the magnetic field through the reversal of the interfacial CrI$_3$ layer, the dominant trion helicity switches (compare the spectra at $B=\SI{0}{T}$ and \SI{-0.4}{T}). Thus, in the AFM domain, the trion response follows the sign of the interfacial proximity exchange field   during magnetization reversal.

The mixed AFM/FM domain (right panel) behaves differently. In the high magnetic field regimes ($B=\SI{\pm2.2}{T}$), where all CrI$_3$ layers are aligned parallel, the spectra closely resemble those of the AFM domain: the trion is again seen in only the one helicity that exhibits the neutral exciton resonance at higher energy. In the intermediate magnetic field regime between \SI{0.3}{T} and \SI{-0.6}{T}, however, where the magnetic layer sequence contains both AFM and FM interlayer coupling, the trion response changes qualitatively. The trion resonance is now visible in both light helicities with comparable but overall reduced amplitudes.

We quantify this behavior by integrating the RC signal over the trion resonance and subtract the background amplitude extracted from a spectrum taken at $B=\SI{2.5}{~T}$ (see methods section for more details). The resulting helicity-resolved amplitudes for the magnetic down-sweep direction are shown in Fig.~\ref{fig:fig3}b. In the  AFM domain, the two helicities show a complementary field dependence and interchange at the magnetic switching field (coercive field) of the interfacial CrI$_3$ layer (see dashed line in the upper panel of Fig.~\ref{fig:fig3}b). By contrast, in the mixed AFM/FM domain between approximately $\SI{-0.9}{T}$ and $\SI{0.5}{T}$ both helicities exhibit trion resonances with comparable amplitudes (see magnetic range between the two dashed lines in the lower panel of Fig.~\ref{fig:fig3}b). This is precisely the magnetic field range with the partial FM ordering (compare to B$_{2}^{*}$ and B$_{3}^{*}$ in Figs.~\ref{fig:fig2}b,d,f).

\subsection{Discussion}

A possible interpretation of our finding is that the mixed AFM/FM configuration modifies the valley-dependent occupation of the WSe$_2$ valence band. As the formation of positive trions requires resident holes, the trion oscillator strength is strongly affected by the relative occupation of the spin-valley-locked valence band states. Both in the AFM domain and at large magnetic fields also in the mixed domain, the helicity selectivity of the trion resonance indicates a large imbalance of the holes between the two valleys. However, at smaller fields in the mixed AFM/FM configuration, the appearance of $X_+$ in both helicities indicates a reduced valley selective hole imbalance. This may originate from a modified valence band alignment, a reduced effective valence band splitting, or changes in interlayer charge transfer. The present experiment does not allow us to separate these contributions, as the measured neutral-exciton splitting remains of comparable splitting for all configurations.  

An alternative explanation could be that the two domains may differ by crystallographic stacking order~\cite{Li2019_pressure_CrI3, Song2019_pressure_CrI3, Soriano2019, Sivadas2018}. Indeed, mixed AFM/FM interlayer coupling in CrI$_3$ has been linked to stacking dependent interlayer exchange. However, stacking alone cannot explain the observed full magnetic field dependence. The two domains show their strongest differences only at smaller magnetic fields, whereas the trion response becomes the same once the CrI$_3$ trilayer is driven into a fully parallel alignment at high fields. As the stacking order is not expected to change during the magnetic field sweep, the disappearance of the domain-dependent response at high magnetic fields indicates that the relevant control parameter is the magnetic layer sequence itself.

\section {Conclusion}
In conclusion, we have investigated magnetic proximity effects in a gate-tunable WSe$_2$/trilayer-CrI$_3$ heterostructure exhibiting two distinct magnetic domains with purely AFM and mixed AFM/FM interlayer coupling. By combining helicity-resolved reflection-contrast spectroscopy with magnetic circular dichroism, we reconstruct the layer-resolved magnetic configurations and investigate their influence on the neutral- and charged-exciton responses of WSe$_2$. While the sign of the neutral-exciton valley splitting predominantly follows the magnetization of the interfacial CrI$_3$ layer, the positive-trion response exhibits a qualitatively different sensitivity to the magnetic configuration. In particular, the mixed AFM/FM domain shows trion resonances in both optical helicities, whereas the purely AFM domain remains strongly helicity-selective. This difference disappears when the CrI$_3$ trilayer is driven into a fully polarized state.
Our results therefore refine the common picture of magnetic proximity effects in WSe$_2$/CrI$_3$ heterostructures. While the neutral exciton mainly senses the interfacial CrI$_3$ magnetization, the positive trion probes the hole-doped electronic structure and is sensitive to magnetic order beyond the immediate interface. Although the present measurements do not allow us to independently resolve the contributions of individual CrI$_3$ layers to the trion response, they demonstrate that the magnetic configuration of the multilayer stack can influence the optical properties of charged excitations in the adjacent TMD. This distinction is particularly relevant for proximity-engineered spintronic and valleytronic devices, where the relevant degrees of freedom are often resident electrons or holes rather than neutral excitons.
 
\section{Methods}
\subsection{Sample preparation}
All 2D-materials were mechanically exfoliated from bulk crystals onto Si/SiO$_2$ substrates using heat-assisted exfoliation. Monolayer WSe$_2$ and trilayer CrI$_3$ were exfoliated inside a nitrogen-filled glovebox, while graphite and hBN were exfoliated under ambient conditions and subsequently transferred into the glovebox. Suitable flakes were identified using an automated flake-detection algorithm~\cite{Uslu2024,Uslu2025Dec}. The heterostructure was assembled inside the glovebox using a polycarbonate-assisted dry-transfer method~\cite{Pizzocchero2016} with a droplet-shaped tool~\cite{Bisswanger2022Jun}. 

Special care was taken to minimize degradation of CrI$_3$, which is sensitive to oxygen, water, and UV light~\cite{Shcherbakov2018}. To reduce high-energy light exposure, long-pass filters (\SI{530}{nm} cut-off wavelength) were installed in the used microscopes, and the glovebox was equipped with UV-blocking foil (\SI{520}{nm} cut-off wavelength). Substrates were heated to \SI{325}{\degree C} and subsequently cooled to \SI{70}{\degree C} before exfoliation to reduce adsorbed water and surface contamination. The trilayer CrI$_3$ used in this work was encapsulated within \SI{28}{h} after exfoliation. After hBN encapsulation, the heterostructure remained stable under ambient conditions.
Electrical contacts were defined by electron-beam lithography. The contact regions were opened by SF$_6$/Ar reactive-ion etching, followed by Cr/Au metallization.

The two magnetic domains cannot intentionally be fabricated. While the CrI$_3$ layer number was determined before assembly, its local crystallographic stacking order was not characterized.

\subsection{Helicity-resolved optical reflectance measurements}
Optical reflectance measurements were performed in a liquid-helium bath cryostat (attoLIQUID1000, attocube systems AG) equipped with an optical head, a spectrometer, and a superconducting magnet immersed in the helium bath, allowing the application of out-of-plane magnetic fields.
A tungsten–halogen white-light source is collimated and directed through a linear polarizer, followed by a liquid-crystal variable retarder, which is used to switch between $\sigma^{+}$ and $\sigma^{-}$ circular polarization. The circularly polarized light is routed via a 50:50 beamsplitter and focused onto the sample using a 100× objective, resulting in a spot diameter of approximately \SI{0.5}{\micro \m} full-width-half-maximum.
The reflected light retraces the same optical path through the objective and the 50:50 beam splitter, passes an additional 90:10 beam splitter, and is subsequently coupled into an optical fiber connected to a spectrometer equipped with a 1200 grooves/mm grating. For each spatial map, reflected intensities are recorded sequentially using $\sigma^{+}$ and $\sigma^{-}$ circular polarization. From these helicity-resolved reflectance signals, both the reflectance contrast (RC) and the magnetic circular dichroism (MCD) are extracted, as defined below.

The RC is defined as: 
\begin{equation*}
RC(E) = \frac{I_{\mathrm{s}}(E) - I_{\mathrm{bg}}(E)}
{I_{\mathrm{ref}}(E) - I_{\mathrm{bg}}(E)} ,
\label{eq:RC}
\end{equation*}
where $I_{\mathrm{s}}(E)$ denotes the reflected intensity measured on the sample region,
$I_{\mathrm{ref}}(E)$ the reflected intensity measured on a reference region without WSe$_2$,
and $I_{\mathrm{bg}}(E)$ the background signal.

The MCD is defined as: 
\begin{equation*}
\mathrm{MCD}(E) = 
\frac{I^{+}(E) - I^{-}(E)}{I^{+}(E) + I^{-}(E)} ,
\label{eq:MCD}
\end{equation*}
with $I^{+}(E)$ and $I^{-}(E)$ as the reflected intensities measured with
$\sigma^{+}$ and $\sigma^{-}$ circular polarization, respectively.
The MCD signal of CrI$_3$ is obtained by integrating the helicity-resolved
intensities over the energy range from \SI{1.743}{\eV} to
\SI{1.755}{\eV}.
The resulting MCD is calculated separately for the reference and both sample regions (I and II), averaged within each region, and the reference contribution is subtracted from the signals of both samples regions.

\subsection{Exciton energy and valley splitting determined by Fano fits}
To determine the energy of the neutral exciton, we fit the helicity-resolved RC spectra with Fano resonances. The RC spectrum is modeled as the sum of two Fano resonances, describing the neutral exciton X$_{0}$ and the positively charged trion X$_{+}$, together with a constant offset,
\begin{equation}
F(E)=f_{X^{0}}(E)+f_{X^{+}}(E)+b_0 .
\end{equation}

Each resonance is described by
\begin{equation}
f(E)=\frac{A_0}{1+q^2}\frac{\left[\Gamma q+(E-E_0)\right]^2}{(E-E_0)^2+\Gamma^2}.
\end{equation}

Here, $E_0$ is the resonance energy, $\Gamma$ the linewidth, $A_0$ the amplitude, $q$ the Fano asymmetry parameter, and $b_0$ a constant background offset. The Fano parameter $q$ is taken to be identical for the exciton and trion contributions within a given spectrum.

For each magnetic field and circular helicity, the RC spectrum is fitted independently. The fits are restricted to the spectral range containing the neutral exciton and trion resonances. The exciton valley splitting is then obtained from the difference between the fitted exciton resonance energies in $\sigma^{+}$ and $\sigma^{-}$ polarization,
\begin{equation}
\Delta E = E_{X^{0}}^{\sigma^{+}} - E_{X^{0}}^{\sigma^{-}}.
\end{equation}

\subsection{Calculation of trion amplitude}
The helicity-resolved trion amplitude was extracted by integrating the RC signal over a fixed spectral window covering the $X_{+}$ resonance. For both  the  AFM and mixed AFM/FM  domain we used the energy interval
\begin{equation}
\Delta E_{\mathrm{trion}}
=
[\SI{1.682}{eV},\SI{1.696}{eV}] .
\end{equation}
For each helicity $\sigma$ and magnetic field $B$, we calculate the integrated RC signal
\begin{equation}
S^{\sigma}(B)=\sum_{E \in \Delta E_{\mathrm{trion}}} RC^{\sigma}(B,E).
\end{equation}
As the trion appears as a dip in the RC spectra, a decrease of  $S_{\mathrm{ref}}^{\sigma}$ corresponds to an increase of the trion oscillator strength. To define a positive trion amplitude, we subtract the integrated values from a helicity and domain-specific value 
\begin{equation}
A_{\mathrm{trion}}^{\sigma}(B)=S_{\mathrm{ref}}^{\sigma}-S_{\mathrm{trion}}^{\sigma}(B).
\end{equation}
Here, $S_{\mathrm{trion}}^{\sigma}(B)$ was obtained from a high-field spectrum in which the trion contribution in the corresponding helicity was minimal. With this convention, a deeper trion dip yields a larger positive value of $A_{\mathrm{trion}}^{\sigma}(B)$.

\textbf{Acknowledgements} 
This work was supported by the Deutsche Forschungsgemeinschaft (DFG, German Research Foundation) - 576172609 and by the Deutsche Forschungsgemeinschaft (DFG, German Research Foundation) under Germany's Excellence Strategy – Cluster of Excellence Matter and Light for Quantum Computing (ML4Q) EXC 2004/2 – 390534769. K.W. and T.T. acknowledge support from the JSPS KAKENHI (Grant Numbers 21H05233 and 23H02052), the CREST
(JPMJCR24A5), JST and World Premier International
Research Center Initiative (WPI), MEXT, Japan.

\textbf{Data availability}
The data supporting the findings are available in a Zenodo repository under accession code https://doi.org/10.5281/zenodo.21821738.

\end{document}